\documentclass[11pt]{article}
\usepackage{eqsection,subeqnarray,indent,amsfonts}

\begin{document}

\def\thefootnote{\fnsymbol{footnote}}   

\begin{center}

\LARGE\bf Chromatic Polynomials, Potts Models \\
\LARGE\bf      and All That \\[5mm]
\large\rm Alan D. Sokal\footnote{Supported in part by NSF grants PHY--9520978
    and PHY--9900769.\break\hfill
    {\em E-mail address:} sokal@nyu.edu} \\[2mm]
\small\it Department of Physics, New York University,
    New York, NY 10003, USA \\[8mm]
\end{center}

\def\thefootnote{\arabic{footnote}}   
\setcounter{footnote}{0}              
 
\begin{abstract}
The $q$-state Potts model can be defined on an arbitrary finite graph,
and its partition function encodes much important information
about that graph, including its chromatic polynomial, flow polynomial
and reliability polynomial.
The complex zeros of the Potts partition function
are of interest both to statistical mechanicians
and to combinatorists.
I give a pedagogical introduction to all these problems,
and then sketch two recent results:
(a) Construction of a countable family of planar graphs whose
chromatic zeros are dense in the whole complex $q$-plane
except possibly for the disc $|q-1| < 1$.
(b) Proof of a universal upper bound on
the $q$-plane zeros of the chromatic polynomial
(or antiferromagnetic Potts-model partition function)
in terms of the graph's maximum degree.
\end{abstract}

\newcommand{\be}{\begin{equation}}
\newcommand{\ee}{\end{equation}}
\newcommand{\<}{\langle}
\renewcommand{\>}{\rangle}
\newcommand{\widebar}{\overline}
\def\reff#1{(\protect\ref{#1})}
\def\spose#1{\hbox to 0pt{#1\hss}}
\def\ltapprox{\mathrel{\spose{\lower 3pt\hbox{$\mathchar"218$}}
 \raise 2.0pt\hbox{$\mathchar"13C$}}}
\def\gtapprox{\mathrel{\spose{\lower 3pt\hbox{$\mathchar"218$}}
 \raise 2.0pt\hbox{$\mathchar"13E$}}}
\def\textprime{${}^\prime$}
\def\proof{\par\medskip\noindent{\sc Proof.\ }}
\def\qed{\hbox{\hskip 6pt\vrule width6pt height7pt depth1pt \hskip1pt}\bigskip}
\def\proofof#1{\bigskip\noindent{\sc Proof of #1.\ }}
\def\half{ {1 \over 2} }
\def\third{ {1 \over 3} }
\def\twothird{ {2 \over 3} }
\def\smfrac#1#2{\textstyle{#1\over #2}}
\def\smhalf{ \smfrac{1}{2} }
\newcommand{\real}{\mathop{\rm Re}\nolimits}
\renewcommand{\Re}{\mathop{\rm Re}\nolimits}
\newcommand{\imag}{\mathop{\rm Im}\nolimits}
\renewcommand{\Im}{\mathop{\rm Im}\nolimits}
\newcommand{\sgn}{\mathop{\rm sgn}\nolimits}
\def\hboxscript#1{ {\hbox{\scriptsize\em #1}} }

\def\scra{\mathcal{A}}
\def\scrc{\mathcal{C}}
\def\scrf{\mathcal{F}}
\def\scrg{\mathcal{G}}
\def\scrl{\mathcal{L}}
\def\scro{\mathcal{O}}
\def\scrp{\mathcal{P}}
\def\scrr{\mathcal{R}}
\def\scrs{\mathcal{S}}
\def\scrt{\mathcal{T}}
\def\scrv{\mathcal{V}}
\def\scrz{\mathcal{Z}}

\def\q{{\sf q}}

\def\Z{{\mathbb Z}}
\def\R{{\mathbb R}}
\def\C{{\mathbb C}}
 
\newtheorem{theorem}{Theorem}[section]
\newtheorem{proposition}[theorem]{Proposition}
\newtheorem{lemma}[theorem]{Lemma}
\newtheorem{corollary}[theorem]{Corollary}


\newenvironment{sarray}{
          \textfont0=\scriptfont0
          \scriptfont0=\scriptscriptfont0
          \textfont1=\scriptfont1
          \scriptfont1=\scriptscriptfont1
          \textfont2=\scriptfont2
          \scriptfont2=\scriptscriptfont2
          \textfont3=\scriptfont3
          \scriptfont3=\scriptscriptfont3
        \renewcommand{\arraystretch}{0.7}
        \begin{array}{l}}{\end{array}}
 
\newenvironment{scarray}{
          \textfont0=\scriptfont0
          \scriptfont0=\scriptscriptfont0
          \textfont1=\scriptfont1
          \scriptfont1=\scriptscriptfont1
          \textfont2=\scriptfont2
          \scriptfont2=\scriptscriptfont2
          \textfont3=\scriptfont3
          \scriptfont3=\scriptscriptfont3
        \renewcommand{\arraystretch}{0.7}
        \begin{array}{c}}{\end{array}}

\section{Introduction}   \label{sec1}

The Potts model \cite{Potts_52,Wu_82,Wu_84}
plays an important role in the general theory of critical phenomena,
especially in two dimensions
\cite{Baxter_82,Itzykson_collection,DiFrancesco_97},
and has applications to various condensed-matter systems \cite{Wu_82}.
Ferromagnetic Potts models have been extensively studied over the
last two decades, and much is known about their
phase diagrams \cite{Wu_82,Wu_84}
and critical exponents \cite{Itzykson_collection,DiFrancesco_97,Nienhuis_84}.
But for antiferromagnetic Potts models, many basic questions remain open:
Is there a phase transition at finite temperature, and if so, of what order?
What is the nature of the low-temperature phase(s)?
If there is a critical point, what are the critical exponents and the
universality classes?
The answers to these questions are expected to be highly lattice-dependent,
in sharp contrast to the universality typically enjoyed by ferromagnets.
This suggests studying antiferromagnetic Potts models on general graphs,
in order to learn {\em which}\/ combinatorial properties of the lattice
correlate with which properties of the phase diagram.

So let $G = (V,E)$ be a finite undirected graph
with vertex set $V$ and edge set $E$.
For each positive integer $q$,
let $P_G(q)$ be the number of ways that the vertices of $G$
can be assigned ``colors'' from the set $\{ 1,2,\ldots,q \}$
in such a way that adjacent vertices always receive different colors.
It is not hard to show (see below) that $P_G(q)$ is the restriction
to $\Z_+$ of a polynomial in $q$.
This (obviously unique) polynomial is called the
{\bf chromatic polynomial}\/ of $G$,
and can be taken as the {\em definition}\/ of $P_G(q)$ for
arbitrary real or complex values of $q$.\footnote{
   See \cite{Read_68,Read_88} for excellent reviews on chromatic polynomials,
   and \cite{Chia_97} for an extensive bibliography.
}

The chromatic polynomial was introduced in 1912 by Birkhoff \cite{Birkhoff_12}.
The original hope was that study of the real or complex zeros of $P_G(q)$
might lead to an analytic proof of the Four-Color Conjecture,
which states that $P_G(4) > 0$ for all loopless planar graphs $G$.
To date this hope has not been realized, although combinatoric proofs
of the Four-Color Theorem have been found \cite{Appel_89,Robertson_97}.
Even so, the zeros of $P_G(q)$ are interesting in their own right
and have been extensively studied.
Most of the available theorems concern real zeros,
but there has been some study (mostly numerical) of complex zeros as well,
by both mathematicians
and physicists
(see \cite{Sokal_99} for an extensive list of references).

A more general polynomial can be obtained as follows:
Assign to each edge $e \in E$ a real or complex weight $v_e$.
Then define
\be
   Z_G(q, \{v_e\})   \;=\;
   \sum_{ \{\sigma_x\} }  \,  \prod_{e \in E}  \,
      \biggl[ 1 + v_e \delta(\sigma_{x_1(e)}, \sigma_{x_2(e)}) \biggr]
   \;,
 \label{eq1.1}
\ee
where the sum runs over all maps $\sigma\colon\, V \to \{ 1,2,\ldots,q \}$,
the $\delta$ is the Kronecker delta,
and $x_1(e), x_2(e) \in V$ are the two endpoints of the edge $e$.
It is not hard to show (see below)
that $Z_G(q, \{v_e\})$ is the restriction to $q \in \Z_+$ of
a polynomial in $q$ and $\{v_e\}$.
If we take $v_e = -1$ for all $e$, this reduces to the chromatic polynomial.
If we take $v_e = v$ for all $e$, this defines a two-variable polynomial
$Z_G(q,v)$ that was introduced
implicitly by Whitney \cite{Whitney_32a,Whitney_32b,Whitney_33}
and explicitly by Tutte \cite{Tutte_47,Tutte_54};
it is known variously (modulo trivial changes of variable)
as the {\bf dichromatic polynomial}\/, the {\bf dichromate}\/,
the {\bf Whitney rank function}\/ or the {\bf Tutte polynomial}\/
\cite{Welsh_93,Biggs_93}.

In statistical mechanics, \reff{eq1.1} is the partition function of the
{\bf \mbox{\boldmath $q$}-state Potts model}\/ \cite{Potts_52,Wu_82,Wu_84}
with Hamiltonian
\be
   H(\{\sigma_x\})   \;=\;
   - \sum_{e \in E} J_e \delta(\sigma_{x_1(e)}, \sigma_{x_2(e)})
   \;;
\ee
here we have written
\be
   v_e  \;=\;  e^{\beta J_e} - 1  \;.
\ee
A coupling $J_e$ (or $v_e$)
is called {\bf ferromagnetic}\/ if $J_e \ge 0$ ($v_e \ge 0$)
and {\bf antiferromagnetic}\/ if $-\infty \le J_e \le 0$ ($-1 \le v_e \le 0$).
The chromatic polynomial thus corresponds to the
zero-temperature limit of the antiferromagnetic Potts model.

To see that $Z_G(q, \{v_e\})$ is indeed a polynomial in its arguments
(with coefficients that are in fact 0 or 1), we proceed as follows:
In \reff{eq1.1}, expand out the product over $e \in E$,
and let $E' \subset E$ be the set of edges for which the term
$v_e \delta_{\sigma_{x_1(e)}, \sigma_{x_2(e)}}$ is taken.
Now perform the sum over configurations $\{ \sigma_x \}$:
in each connected component of the subgraph $(V,E')$
the spin value $\sigma_x$ must be constant,
and there are no other constraints.
Therefore,
\be
   Z_G(q, \{v_e\})   \;=\;
   \sum_{ E' \subset E }  q^{k(E')}  \prod_{e \in E'}  v_e
   \;,
 \label{eq1.2}
\ee
where $k(E')$ is the number of connected components
(including isolated vertices) in the subgraph $(V,E')$.
The expansion \reff{eq1.2} was discovered by
Birkhoff \cite{Birkhoff_12} and Whitney \cite{Whitney_32a}
for the special case $v_e = -1$ (see also Tutte \cite{Tutte_47,Tutte_54});
in its general form it is due to
Fortuin and Kasteleyn \cite{Kasteleyn_69,Fortuin_72}
(see also \cite{Edwards-Sokal}).
We take \reff{eq1.2} as the {\em definition}\/ of $Z_G(q, \{v_e\})$
for arbitrary complex $q$ and $\{v_e\}$.

Let us note that a vast amount of combinatorial information about the graph $G$
is encoded in $Z_G(q, \{v_e\})$.  Special cases of $Z_G$ include
not only the chromatic polynomial ($v=-1$) but also the flow polynomial
($v=-q$), the reliability polynomial ($q=0$) and several other
quantities \cite{Welsh_93}.

In statistical mechanics, a very important role is played by the
complex zeros of the partition function \cite{Yang-Lee_52}.
Recall that a {\em phase transition}\/ occurs whenever
one or more physical quantities (e.g.\ the energy or the magnetization)
depend nonanalytically on one or more control parameters
(e.g.\ the temperature or the magnetic field).
Now, such nonanalyticity is manifestly impossible in
\reff{eq1.1}/\reff{eq1.2} for any finite graph $G$.
Rather, phase transitions arise only in the {\em infinite-volume limit}\/.
That is, we consider some countably infinite graph
$G_\infty = (V_\infty, E_\infty)$
--- usually a regular lattice, such as $\Z^d$ with nearest-neighbor edges ---
and an increasing sequence of finite subgraphs $G_n = (V_n, E_n)$.
It can then be shown
(under modest hypotheses on the $G_n$)
that the {\bf (limiting) free energy per unit volume}\/
\be
   f_{G_\infty}(q,v)   \;=\;
   \lim_{n \to \infty}   |V_n|^{-1}  \log Z_{G_n}(q,v)
 \label{limiting_free_energy}
\ee
exists for all {\em nondegenerate physical}\/ values
of the parameters,
namely either
\begin{quote}
\begin{itemize}
  \item[(a)]     $q$ integer $\ge 1$ and $-1 < v < \infty$
     \quad  [using \reff{eq1.1}:  see e.g.\ \cite[Section I.2]{Israel_79}]
  \item[or (b)]  $q$ real $> 0$ and $0 \le v < \infty$
     \quad  [using \reff{eq1.2}:  see \cite[Theorem 4.1]{Grimmett_95}
                                  and \cite{Grimmett_78,Seppalainen_98}].
\end{itemize}
\end{quote}
This limit $f_{G_\infty}(q,v)$ is in general a continuous function of $v$;
but it can fail to be a real-analytic function of $v$,
because complex singularities of $\log Z_{G_n}(q,v)$
--- namely, complex zeros of $Z_{G_n}(q,v)$ ---
can approach the real axis in the limit $n \to\infty$.
Therefore, the possible points of physical phase transitions
are precisely the real limit points of such complex zeros \cite{Yang-Lee_52}.
As a result, theorems that constrain the possible location of
complex zeros of the partition function are of great interest.
In particular, theorems guaranteeing that a certain complex domain
is free of zeros are often known as {\em Lee-Yang theorems}\/.


\section{A Lee-Yang Theorem for Chromatic Polynomials?}  \label{sec2}

Let me now review some known facts about the {\em real}\/ zeros of the
chromatic polynomial $P_G(q)$, in order to motivate some conjectures
concerning the {\em complex}\/ zeros:

1)  It is not hard to show that for any loopless graph $G$ with $n$ vertices,
$(-1)^n P_G(q) > 0$ for real $q < 0$ \cite{Read_88}.
It is then natural to ask whether the absence of negative real zeros
might be the tip of the iceberg of a Lee-Yang theorem:
that is, might there exist a {\em complex}\/ domain $D$
containing $(-\infty,0)$ that is zero-free for all $P_G$?
One's first guess is that the half-plane $\Re q < 0$ might be zero-free
\cite{Farrell_80}.
This turns out to be false:
examples are known of loopless graphs $G$ whose chromatic polynomials
have zeros with real part as negative as $\approx -0.7$
\cite{Baxter_87,Read_91,Shrock_98b,Shrock_98e,Shrock_99b,Shrock_99c,%
Salas-Sokal_in_prep}.
Nevertheless, it is not ruled out that some smaller domain
$D \supset (-\infty,0)$ might be zero-free.

2) For any loopless {\em planar}\/ graph $G$,
Birkhoff and Lewis \cite{Birkhoff_46} proved in 1946 that
$P_G(q) > 0$ for real $q \ge 5$;
we now know that $P_G(4) > 0$
\cite{Appel_89,Robertson_97};
and it is very likely (though not yet proven as far as I know)
that $P_G(q) > 0$ also for $4 < q < 5$.
Thus, it is natural to conjecture that might exist
a {\em complex}\/ domain $D$ containing $(4,\infty)$ [or $(5,\infty)$]
that is zero-free for all planar $P_G$.
One's first guess might be that $\Re q > 4$ works.
This again turns out to be false:
examples are known of loopless planar graphs $G$ whose chromatic polynomials
have zeros with real part as large as $\approx 4.2$
\cite{Baxter_87,Salas-Sokal_in_prep}.
Nevertheless, it is not ruled out that some smaller domain
$D \supset (4,\infty)$ might be zero-free.

As with most of my conjectures, these two are false;
but what is interesting is that they are {\em utterly, spectacularly false}\/,
for I can prove:
\begin{theorem}  {\bf \protect\cite{Sokal_hierarchical}}
   \label{thm1}
There is a countably infinite family of planar graphs
whose chromatic zeros are, taken together,
dense in the entire complex $q$-plane
with the possible exception of the disc $|q-1| < 1$.
\end{theorem}
The graphs in question are ``theta graphs'' $\Theta_{s,p}$
obtained by parallel-connecting $p$ chains each of which has
$s$ edges in series.
Theorem~\ref{thm1} is in fact a corollary of a more general result
for the two-variable polynomials $Z_G(q,v)$:
\begin{theorem}  {\bf \protect\cite{Sokal_hierarchical}}
   \label{thm2}
Fix complex numbers $q_0,v_0$ satisfying $|v_0| \le |q_0 + v_0|$.
Then, for each $\epsilon > 0$, there exist numbers $s_0 < \infty$
and $p_0(s) < \infty$ such that for all $s \ge s_0$ and $p \ge p_0(s)$:
\begin{itemize}
   \item[(a)]  $Z_{\Theta_{s,p}}(\,\cdot\,, v_0)$ has a zero in the disc
      $|q - q_0| < \epsilon$.
   \item[(b)]  $Z_{\Theta_{s,p}}(q_0, \,\cdot\,)$ has a zero in the disc
      $|v - v_0| < \epsilon$.
\end{itemize}
\end{theorem}
(Setting $v_0 = -1$, Theorem~\ref{thm2}(a) implies Theorem~\ref{thm1}.)

The intuition behind Theorem~\ref{thm2} is based on recalling
the rules for parallel and series combination of Potts edges:
\begin{itemize}
   \item[\quad] {\bf Parallel}:
      $v_{\hboxscript{eff}} = v_1 + v_2 + v_1 v_2$
      \qquad\, (mnemonic: $1 + v$ multiplies)
   \item[\quad] {\bf Series}:
      $v_{\hboxscript{eff}} = v_1 v_2 / (q + v_1 + v_2)$
      \qquad (mnemonic: $v/(q+v)$ multiplies)
\end{itemize}
In particular, if $0 < |v/(q+v)| < 1$, then putting a large number $s$ of edges
in series drives the effective coupling $v_{\hboxscript{eff}}$
to a small (but nonzero) number;
moreover, by small perturbations of $v$ and/or $q$
we can give $v_{\hboxscript{eff}}$ any phase we please.
But then, by putting a large number $p$ of such chains in parallel,
we can make the resulting $v_{\hboxscript{eff}}$ lie anywhere
in the complex plane we please.
In particular, we can make $v_{\hboxscript{eff}}$ equal to $-q$,
which causes the partition function $Z_{\Theta_{s,p}}$ to be zero.

The proof of Theorem~\ref{thm2} is based on a complex-variables result due to
Beraha, Kahane and Weiss \cite{BKW_75,BKW_78,Beraha_79,Beraha_80};
see \cite{Sokal_hierarchical} for details.

\section{A Bound in Terms of Maximum Degree}  \label{sec3}

Very little is known rigorously about the phase diagrams
of Potts antiferromagnets, but one general result is available:
for $q$ large enough (how large depends on the lattice in question),
the antiferromagnetic $q$-state Potts model has
a unique infinite-volume Gibbs measure and
exponential decay of correlations
at all temperatures, {\em including zero temperature}\/.
(Physically, the system is disordered as a result of the large
ground-state entropy,
so that zero temperature belongs to the high-temperature regime!)
More precisely,
using the Dobrushin uniqueness theorem \cite{Georgii_88,Simon_93}
it can be proven \cite{Salas_97}
that for a countable graph $G$ in which every vertex has
at most $r$ nearest neighbors,
the $q$-state Potts-model Gibbs measure on $G$ is unique
for all integer $q > 2 r$ whenever
$-1 \le v_e \le 0$ for all edges $e$.

Uniqueness of the Gibbs measure means that
{\em first-order}\/ phase transitions are excluded,
but higher-order phase transitions are not necessarily ruled out.
In particular, the foregoing result does not imply the analyticity
of the free energy for large-$q$ Potts antiferromagnets,
but it does make it plausible.
Of course, a result that holds for {\em integer}\/ $q > q_0$
need not hold for all {\em real}\/ $q > q_0$, much less for a complex
neighborhood of that real semi-axis; but it does suggest that such a
result might be true.
This led me to conjecture that there might exist universal constants
$C(r) < \infty$ such that, for all loopless graphs $G$ of
maximum degree $\le r$,
the zeros of the chromatic polynomial $P_G(q)$
[and more generally of antiferromagnetic Potts partition functions]
lie in the disc $|q| < C(r)$.\footnote{
   I later learned that this result for $P_G(q)$ had been conjectured
   by mathematicians more than 25 years ago
   \protect\cite{Biggs_72} \protect\cite[Question 6.1]{Brenti_94}.
}
This is in fact the case, and the bound holds
throughout the ``complex antiferromagnetic regime'' $|1 + v_e| \le 1$:

\begin{theorem} {\bf \protect\cite{Sokal_99}}
  \label{thm3.1}
There exist universal constants $C(r) \le 7.963907 r$ such that,
for all loopless graphs $G=(V,E)$  of maximum degree $\le r$,
equipped with complex edge weights $\{ v_e \}_{e \in E}$
satisfying $|1 + v_e| \le 1$ for all $e$,
the zeros of $Z_G(q, \{v_e\})$ all lie in the disc $|q| < C(r) v_{max}$,
where $v_{max} = \max\limits_{e \in E} |v_e|$.
In particular, the zeros of $P_G(q)$ all lie in the disc $|q| < C(r)$.
\end{theorem}
This linear dependence on $r$ is best possible,
as the example of the complete graph $K_{r+1}$ shows that $C(r) \ge r$.
However, the constant 7.963907 can presumably be improved
(see Section \ref{sec4} below).

It is amusing to note that the presence of {\em one}\/ vertex
of large degree cannot lead to large chromatic roots.
More precisely, if all but one of the vertices of $G$ have degree $\le r$,
then I can prove that
the zeros of $P_G(q)$ lie in the disc $|q| < C(r) + 1$ \cite{Sokal_99}.
Please note that a result of this kind {\em cannot}\/ hold if ``all but one''
is replaced by ``all but two'', for in this case the chromatic roots
can be unbounded, even when $r=2$ and $G$ is planar,
as the graphs $\Theta_{s,p}$ show.

The proofs of these results are based on well-known methods
of mathematical statistical mechanics.
The first step is to transform the
Whitney--Tutte--Fortuin--Kasteleyn representation \reff{eq1.2}
into a gas of ``polymers'' interacting via a hard-core exclusion.
One then invokes the Dobrushin condition \cite{Dobrushin_96a,Dobrushin_96b}
(or the closely related Koteck\'y--Preiss condition 
 \cite{Kotecky_86,Sokal_Mayer_in_prep})
for the nonvanishing of a polymer-model partition function.
Lastly, one verifies these conditions for our particular polymer model,
using a series of simple combinatorial lemmas.
Details can be found in \cite{Sokal_99}.

With a little more work,
it should be possible to extend the arguments of \cite{Sokal_99}
to prove the existence and analyticity
of the limiting free energy per unit volume \reff{limiting_free_energy}
for suitable regular lattices $G_\infty$
and translation-invariant edge weights $v_e$,
in the same region of complex $q$- and $\{v_e\}$-space
where $Z$ is proven to be nonvanishing uniformly in the volume.
In particular, this would provide a convergent expansion for
the limiting free energy in powers of $1/q$.
However, I have not worked out the details.

\section{Some Conjectures and Open Questions}   \label{sec4}

The bound in Theorem \ref{thm3.1} is, of course, far from sharp,
and it is of some interest to speculate on what the best-possible
results might be.  Let us define
\be
   C_{opt}(r)   \;=\;
   \max\{|q| \colon\;  P_G(q) = 0
       \hbox{ for some loopless graph $G$ of maximum degree $r$} \}
   \;.
\ee
The example of the complete graph $K_{r+1}$ shows that $C_{opt}(r) \ge r$.
It is easy to see that $C_{opt}(1) = 1$ and $C_{opt}(2) = 2$;
and there is some evidence that $C_{opt}(3) = 3$.
But, at least for $r \ge 4$,
$C_{opt}(r)$ must in fact be strictly larger than $r$,
as is shown by numerical computations on the
complete bipartite graph $K_{r,r}$ \cite{Sokal_99}.
Indeed, for large $r$ it seems that $K_{r,r}$ has a chromatic zero
of magnitude approximately $1.5r$.
On the other hand, the Dobrushin uniqueness result quoted in
the preceding section suggests that $C(r) = 2r$ might suffice.


We can pose these questions more generally as follows:
Let $\scrg$ be a class of finite graphs,
and let $\scrv$ be a subset of the complex plane.
Then we can ask about the sets
\begin{eqnarray}
   S_1(\scrg, \scrv)   & = &
      \bigcup_{G \in \scrg} \;
      \bigcup_{v \in \scrv} \;
      \{q \in \C \colon\;  Z_G(q,v) \,=\, 0 \}               \\
   S_2(\scrg, \scrv)   & = &
      \bigcup_{G \in \scrg} \;
      \bigcup_{\{v_e\} \colon\; v_e \in \scrv \; \forall e}
      \{q \in \C \colon\;  Z_G(q, \{v_e\}) \,=\, 0 \}
\end{eqnarray}
Among the interesting cases are
the chromatic polynomials $\scrv = \{-1\}$,
the antiferromagnetic Potts models $\scrv = [-1,0]$,
and the complex antiferromagnetic Potts models
$\scrv = A \equiv \{v \in \C \colon\; |1 + v| \le 1 \}$.
Indeed, one moral of this work is that some questions
concerning chromatic polynomials are most naturally studied
in the more general context of antiferromagnetic or
complex antiferromagnetic Potts models
(with not-necessarily-equal edge weights).
The results of \cite{Sokal_99} show that
the set $S_2(\scrg_r, A)$ is bounded,
where $\scrg_r$ is the set of all loopless graphs of maximum degree $\le r$,
and more generally that the set $S_2(\scrg'_r, A)$ is bounded,
where $\scrg'_r$ is the set of all loopless graphs of second-largest
degree $\le r$.
But it would be interesting to examine in more detail the location
of all these sets in the complex plane, and to prove sharper bounds.

Another direction in which the results of \cite{Sokal_99} could be extended
is by finding a criterion {\em weaker}\/ than bounded maximum degree
(or bounded second-largest degree)
under which the zeros of $P_G(q)$ and $Z_G(q, \{v_e\})$ could be shown
to be bounded.
An interesting idea was suggested very recently by
Shrock and Tsai \cite{Shrock_98e,Shrock_99a},
who studied a variety of families of graphs
and arrived at a conjecture that can be rephrased as follows:
For $G=(V,E)$ and $x,y \in V$, define
\begin{subeqnarray}
   \lambda(x,y)
      & = &  \hbox{max \# of edge-disjoint paths from $x$ to $y$} \\
      & = &  \hbox{min \# of edges separating $x$ from $y$}
\end{subeqnarray}
and
\be
   \Lambda(G)   \;=\;   \max\limits_{x \neq y}  \lambda(x,y)   \;.
\ee
Clearly $\lambda(x,y) \le \min[\deg(x), \deg(y)]$
and hence $\Lambda(G) \le $ second-largest degree of $G$.
Now let $\scrg^\Lambda_r$ be the set of all loopless graphs
with $\Lambda(G) \le r$.
Then the conjecture is that the set $S_2(\scrg^\Lambda_r, \scrv)$
is bounded, where $\scrv = \{-1\}$ or $[-1,0]$ or perhaps even $A$.
This possible connection of chromatic-polynomial and Potts-model problems
with max-flow problems is intriguing.
Note that $\Lambda(G)$ and $\Lambda(G, \{v_e\})$
possess a ``naturalness'' property that
maximum degree and its relatives lack:
namely, for any graph $G$ with blocks (2-connected components)
$G_1,\ldots,G_b$, we have
$\Lambda(G, \{v_e\}) = \max\limits_{1 \le i \le b} \Lambda(G_i, \{v_e\})$.

\end{document}